\documentclass{vgtc}                          

\graphicspath{{figures/}{pictures/}{images/}{./}} 

\usepackage{times}                     

\usepackage{tabu}                      
\usepackage{booktabs}                  
\usepackage{lipsum}                    
\usepackage{mwe}                       

\usepackage{mathptmx}                  

\onlineid{0}

\vgtccategory{Research}

\vgtcinsertpkg

\title{Tides of Memory: Digital Echoes of Netizen Remembrance}

\author{Lingyu Peng\thanks{e-mail: lingyupeng6@163.com}\\ %
     \scriptsize Harbin Institute of Technology \\ %
     \scriptsize Future Design School %
\and Chang Ge\thanks{e-mail: changgecn@gmail.com}\\ %
     \scriptsize Harbin Institute of Technology \\ %
     \scriptsize Future Design School %
\and Liying Long\thanks{e-mail: lly15977427336@163.com}\\ %
     \scriptsize Harbin Institute of Technology \\ %
     \scriptsize Future Design School %
\and Xin Li\thanks{e-mail: li1179327296@163.com}\\ %
      \scriptsize Harbin Institute of Technology \\ %
     \scriptsize Future Design School %
\and Xiao Hu\thanks{e-mail: 1140084087@qq.com}\\ %
     \scriptsize Harbin Institute of Technology \\ %
     \scriptsize Future Design School %
\and Pengda Lu\thanks{e-mail: 867103556@qq.com}\\ %
     \scriptsize Harbin Institute of Technology \\ %
     \scriptsize Future Design School %
\and Qingchuan Li\thanks{e-mail: liqingchuan@hit.edu}\\ %
     \scriptsize Harbin Institute of Technology \\ %
     \scriptsize Future Design School %
\and Jiangyue Wu\thanks{Jiangyue Wu is the corresponding author. She is an assistant professor at the Future Design School, Harbin Institute of Technology. This work was supported in part by Guangdong Philosophy and Social Science Foundation 2025 [GD25YYS17]. e-mail: wu.jiangyue@outlook.com}\\ %
     \scriptsize Harbin Institute of Technology \\ %
     \scriptsize Future Design School %
     }

\teaser{
  \centering
  \includegraphics[width=\linewidth]{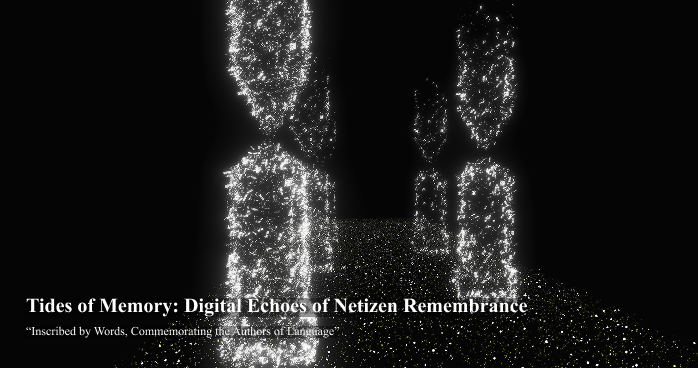}
  \caption{
Tides of Memory: Digital Echoes of Netizen Remembrance}
  \label{fig:teaser}
}

\abstract{
    This artwork presents an interdisciplinary interaction installation that visualizes collective online mourning behavior in China. By focusing on commemorative content posted on Sina Weibo following the deaths of seven prominent Chinese authors, the artwork employs data scraping, natural language processing, and 3D modeling to transform fragmented textual expressions into immersive digital monuments. Through the analysis of word frequencies, topic models, and user engagement metrics, the system constructs a semantic-visual landscape that reflects both authorial legacies and collective memory. This research contributes to the fields of digital humanities, visualization design, and digital memorial architecture by proposing a novel approach for preserving and reactivating collective memory in the digital age. 
} 


\begin{document}


\firstsection{Introduction}

\maketitle

In recent years, celebrities' deaths have sparked widespread online discussions. An increasing number of users engage in spontaneous mourning activities on-line, leading to collective commemorative behaviors \cite{walter2012does}.  In China, the passing of the renowned author Qiong Yao in 2024 triggered massive online mourning, marked by a high-volume hashtag. The Weibo hashtag \textbf{\#QiongYaoPassingAway} reached 860 million views and 355,000 discussions. Users not only recalled her works but also quoted her will as part of their mourning.  This participatory response generated abundant commemorative content and built an emotional support network. Participants changed from passive readers to active creators, reinforcing collective memory through shared emotional expression. This process created strong emotional resonance and collective care \cite{harju2015socially}. This phenomenon also resonates with the identity of the deceased as an author—one who, during their lifetime, crafted narratives and characters that shaped the personal memories of readers. In death, readers used their own words to remember authors, co-creating a rich collective image within a shared emotional framework.

Against this backdrop, the artwork analyzes data from Sina Weibo, a major Chinese social media platform. It focuses on online mourning behaviors following the deaths of Chinese authors since 2016. The artwork collects tens of thousands of mourning posts linked to hashtags for seven authors. These posts are transformed into a three-dimensional monument. This monument creates a tangible commemorative space and visualizes the collective memory and emotions of online users.
\subsection{Contributions}
Existing forms of online mourning are primarily centered on social media platforms, predominantly through commemorative posts under hashtag campaigns. Descriptions of participant groups are largely confined to numerical data. Currently, due to the inherent isolation and virtual nature of cyberspace, interactions among different participants are mostly limited to text and emojis, lacking embodied experiences and making it difficult to form deeper emotional resonance. Furthermore, the linear Browse model of existing platforms leads to rapid coverage of a large volume of mourning content. As a topic's popularity declines, these memories quickly fade on the Internet, lacking an eternal carrier of memory.
The contributions of this paper are as follows:
\begin{itemize}
\item We developed a transformation framework that converts online mourning posts into visible and interactive monuments. This framework also transforms collective hashtag campaigns into immersive memorial spaces, helping people understand how online mourning unfolds.
\item We offer participants new ways of commemoration, extending opportunities for emotional expression. This approach breaks temporal constraints, enabling subsequent individuals to evoke new emotional resonance by visually revisiting past memories, thus joining an ongoing collective memory. This sustains both online mourning and emotional connections to the author.
\item This paper focuses on the phenomenon of online mourning. Through an interdisciplinary research methodology, we visualize its behaviors and outputs, aiming to stimulate subsequent research on this group and promote new forms of "collective care" in the digital age.
\end{itemize}

\subsection{Related work}
\subsubsection{Text Visualization}
The widespread use of online social media has generated vast amounts of textual data. This has driven the advancement of text visualization techniques. Text visualization techniques express complex patterns and semantics through visual encodings, while supporting interactive features that facilitate efficient information retrieval from large-scale datasets \cite{cao2016overview}.This approach turns complex data into engaging narratives. As a result, the information becomes more accessible and easier to understand for a wider audience.

Text visualization typically involves two key stages. The first is text analysis, which converts raw, unstructured text into structured, quantifiable data. The second is visual representation, which transforms the data into visual forms that users can interpret \cite{tang2013survey}. Various methods are available for text analysis. Among them, word frequency analysis, sentiment analysis, and topic modeling are the most common in text visualization \cite{liu2018bridging}. This artwork mainly uses word frequency analysis and topic modeling to examine user-generated comments. It also applies a design strategy that converts two-dimensional data into three-dimensional visual forms. This 2D-to-3D transformation invites users to explore the narratives within the data. It aims to deepen engagement with the emotional and thematic layers of collective online expression.
\subsubsection{Commemorative Space Design}
Memorial spaces are vital carriers of collective memory, weaving emotion and meaning throughout the public realm. From a sociological standpoint, these spaces mobilize feelings, evoke resonance, and help shape shared social recollections \cite{XSYK201207021,1013044470.nh,MYSS201012019} . As Pierre Nora notes, ‘the less memory is experienced from within, the more it relies on external scaffolding and outward signs’ \cite{nora1989between}. Yet traditional monuments and static sculptures often present history in a unidirectional, fixed manner, externalizing memory through cold, hard materials. This approach not only risks fostering passive reception or even alienation among viewers but also leads to a proliferation of commemorative objects without engendering genuine historical understanding \cite{young1999memory}.

\begin{figure}[tb]
 \centering 
 \includegraphics[width=\columnwidth]{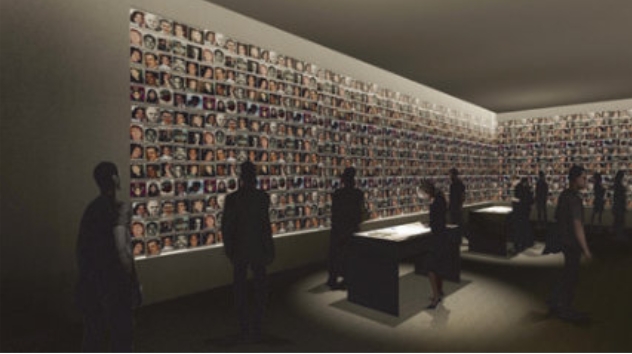}
 \caption{The National September 11 Memorial Museum}
 \label{fig:0.1}
 
\end{figure}

\begin{figure}[tb]
 \centering 
 \includegraphics[width=\columnwidth]{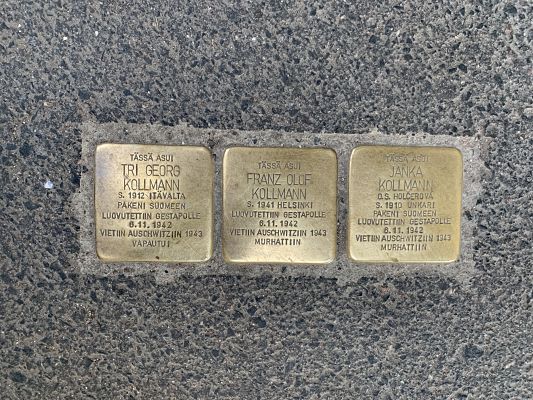}
 \caption{the "Stolpersteine" project by Gunter Demnig}
 \label{fig:0.2}
 
\end{figure}

In response, many international memorial initiatives engage the public directly. The National September 11 Memorial and Museum in New York now features interactive exhibitions and digital archives(see \cref{fig:0.1}). Visitors can hear survivors’ testimonies, examine personal artifacts, and submit or search individual remembrances in real time. This approach turns attendees from passive observers into active co‑creators of collective history \cite{greenwald2010passion}. Similarly, German artist Gunter Demnig’s \textit{Stolpersteine} embeds small brass plaques into sidewalks in front of the victims’ last freely chosen residences (see \cref{fig:0.2}). As Europe’s largest decentralized memorial effort, it anchors remembrance in everyday urban life \cite{mandel2020lieux}. Though these markers lack monumental scale and sometimes take unconventional forms, they subvert grandiose monument conventions and exemplify grassroots memory activism.

\subsection{Framework}
\begin{figure}[tb]
 \centering 
 \includegraphics[width=\columnwidth]{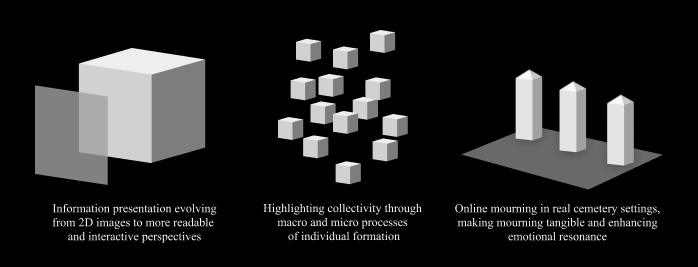}
 \caption{The three-dimensional perspective of information presentation and collective expression in digital tombstones.}
 \label{fig:1.3}
 
\end{figure}

The artwork proposes a systematic design framework for the acquisition, processing, and visualization of online mourning data. We began by collecting user-generated content from Sina Weibo under obituary-related hashtags associated with the deaths of selected authors. A custom-designed web crawler was used to extract mourning posts, which constituted the core dataset. To ensure data reliability, the raw content underwent rigorous preprocessing and analytical refinement.
Building on this foundation, we developed a mapping mechanism to translate the processed data into visually tangible and poetically resonant digital monuments (see \cref{fig:1.3}). This approach mitigates perceptual bias caused by large-scale data disparities while preserving the semantic richness of the commemorative expressions. Finally, we implemented the full digital memorial system using interactive development tools and deployed it as a screen-based immersive experience. This setup invites users to engage emotionally and reflectively, fostering deeper resonance and the continued circulation of collective memory.

\section{Artwork concept}
\subsection{Visualizing Collective Online Mourning}
Online mourning refers to commemorative content posted by the deceased's close relatives, friends, or the general public on digital platforms such as blogs and social media \cite{lingel2013digital}. The affordances of digital technologies provide new avenues for the public to express grief, memorialize the deceased, and offer mutual support. The anonymity and mediation of online media reduce the psychological barriers found in traditional mourning. This allows people to express their emotions more freely and authentically. Thus, online mourning reflects personal memories of the deceased and sincere emotional responses from the public.

In China, online mourning often takes place within specific hashtags. A hashtag is a word or phrase prefixed with the “\#” symbol \cite{wang2016hashtags}, which is a technical feature provided by Weibo. Originally designed as “channel tags” to facilitate user participation in topic-based discussions, hashtags have since garnered scholarly attention due to their significant role in the circulation of social issues \cite{bruns2011use}. Hashtags act as discursive markers. They connect users to conversations about specific events and merge scattered posts into a unified public discourse \cite{highfield2018emoji}. In the context of online mourning, hashtags enable strong emotional connections. They bring together mourners across geographic and cultural boundaries in a shared emotional space. Through collective storytelling, hashtags expand the range and depth of social memory \cite{van2019there}. Hashtag-based commemoration spreads information efficiently. More importantly, it builds a new form of collective care by turning personal grief into shared emotional resonance.

Currently, online mourning appears mainly in two forms: fixed digital spaces, such as memorial webpages, and dynamic spaces, such as mourning posts and hashtags \cite{sofka2009adolescents}. However, the spatial fragmentation inherent in these formats often hinders the formation of meaningful emotional connections among individuals, making it difficult to foster deeper collective resonance. The linear browsing experience further restricts users’ access to the full range of content. In addition, the ephemeral nature of digital platforms leads to the rapid disappearance of mourning-related content. Therefore, the primary objective of our artwork is to explore how to aggregate these fragmented expressions into a tangible entity that can be perceived and understood, and that can generate powerful emotional resonance to influence a broader audience.

\subsection{Commemorating Authors Through Text}
\begin{figure}[tb]
 \centering 
 \includegraphics[width=\columnwidth]{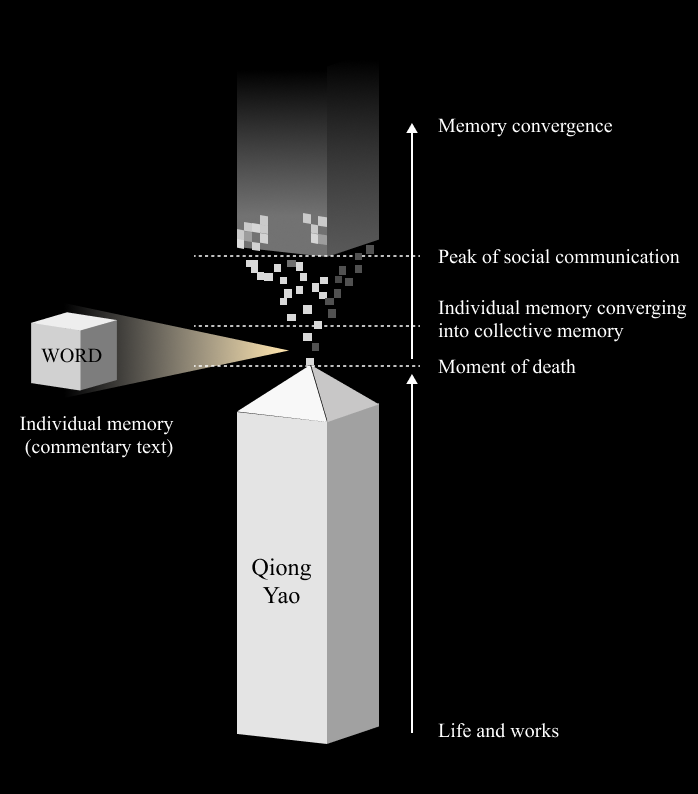}
 \caption{The process of text gathering into a monument}
 \label{fig:2.2}
 
\end{figure}

The main goal of this artwork is to turn widespread digital mourning for deceased authors into a visible and tangible spatial experience. Online users primarily express grief through written language. These texts often go beyond mourning the author's death. They also reflect on the author’s personal impact and recall fictional characters and stories that once shaped users’ lives. In this process, the role of the author changes. Once a creator of stories, the author becomes a subject written about by countless internet users. In response, we propose an artistic framework that uses text to interpret the life and legacy of the author. This creates a dialogic model where users’ words interact with the author’s own words, forming a multifaceted portrait. This approach is inspired by the film \textit{Coco}, which conveys the idea that the act of remembrance sustains the presence of the deceased. Guided by this idea, we aim to preserve the emotional resonance inspired by these authors. At the same time, we want to prevent such sentiments from being fixed in a single moment. Therefore, after visiting the commemorative space, new users are encouraged to add their own words of mourning. These contributions are integrated into the system as evolving data (see \cref{fig:2.2}). In this way, the memorial continues to grow and change, offering a lasting tribute to the authors and affirming their enduring influence.

\section{Implementation}
\subsection{Data Collection and Analytical Framework}
\subsubsection{Data Sources}
The data utilized in this artwork originate from Weibo, a major Chinese social media platform characterized by its rich and diverse user-generated content, including public discourse and commemorative activities related to deceased contemporary Chinese authors. We selected a representative sample of 7 authors spanning different literary generations and stylistic orientations to ensure both breadth and analytical depth. Two primary dimensions guided the data collection: first, authors’ lifetime productivity, measured by the number of published works; second, public responses on social media within the three-month period following each author’s death, captured through four key metrics—Reading Volume, Discussion Volume, Interaction Volume, and Originality Volume (see \cref{tab:vis_papers}).
The observation period was set to three months to balance timeliness with analytical richness. A shorter window would miss secondary attention waves \cite{lehmann2012dynamical,yang2011patterns}, while a longer one risks diminishing returns and topic drift \cite{gama2014survey,knights2009detecting}. In the Chinese cultural context, this period also aligns with significant commemorative milestones such as the 49th and 100th days after death, which often prompt renewed public remembrance \cite{watson1988death}.

\begin{table}[tb]
  \caption{Original data related to mourning posts for the authors.}
  \label{tab:vis_papers}
  \scriptsize%
	\centering%
  \begin{tabu}{%
	r%
	*{7}{c}%
	*{2}{r}%
	}
  \toprule
   Name & \rotatebox{90}{Reading Volume (ten thousand)} & \rotatebox{90}{Discussion Volume ten thousand)} & \rotatebox{90}{Interaction Volume (ten thousand)} & \rotatebox{90}{Originality Volume (ten thousand)} & Time   \\
  \midrule
	Qiong Yao & 86000 & 35.5 & 193.8 & 6.7 & 2024.12 \\
    Qi Bangyuan & 16000 & 2.1 & 7.1 & 0.2414 & 2024.03 \\
    Yang Yi & 3786.4 & 0.8175 & 6.2 & 0.0483 & 2023.01 \\
    Hu Xudong & 1453.2 & 0.2873 & 0.3426 & 0.0194 & 2021.08 \\
    Jin Yong & 210000 & 140.7 & 167.6 & 15 & 2018.10 \\
    Huang Yi & 1510.3 & 0.8072 & 0.9615 & 0.0816 & 2017.04 \\
    Yang Jiang & 14000 & 14.9 & 20.5 & 3.9 & 2016.05 \\
  
  \bottomrule
  \end{tabu}%
\end{table}

To enhance comparability across cases, we adopted a hybrid sampling strategy that combined both temporal recency and engagement intensity. Within the three-month period, salience for each post $i$ was defined as:
\begin{equation}
S_i = E_i \cdot 2^{-\Delta t_i / h}
\end{equation}
where $\Delta t_i$ is the number of days since publication and $h=14$ days, reflecting the exponential decay of online attention \cite{kwak2010twitter, leskovec2009meme,wu2007novelty,yang2011patterns}. Engagement was calculated as:
\begin{equation}
E_i = w_R \tilde{R}_i + w_C \tilde{C}_i + w_L \tilde{L}_i
\end{equation}
where $\tilde{R}_i, \tilde{C}_i, \tilde{L}_i$ are the $z$-score standardized values of $\log(1+x)$ repost, comment, and like counts. Weights $(w_R, w_C, w_L) = (0.40, 0.35, 0.25)$ reflect their respective roles in diffusion, conversational depth, and affective endorsement \cite{gerlitz2013like,zhang2016creates}. Posts above the 70th percentile in $S_i$ were retained.

This approach ensures that the dataset captures not only the most temporally relevant but also the most socially resonant commemorative expressions. By integrating both time sensitivity and public attention, this strategy balances representativeness and salience, while effectively mitigating the distortions introduced by long-term data accumulation or low-engagement content.
\subsubsection{Social Media Data Mining}
A hybrid data collection approach was adopted, integrating Scrapy (v2.7) for structured data extraction and Selenium (v4.x) for interacting with dynamically rendered web content. We initiated the scraping process by constructing compound search queries using the author’s name along with commemoration-related keywords (e.g., "Author Name + passed away", "Author Name + in memory").

For each retrieved Weibo post, the crawler extracted numerical indicators including the number of reads, reposts, comments, and whether the content was original. To ensure data quality, we implemented a multi-step filtering protocol to exclude spam, commercial advertisements, non-textual elements, and irrelevant content. The final curated dataset comprises approximately 8400 valid posts, forming the empirical foundation of the subsequent analysis.

\subsubsection{Data Standardization}
Given the substantial variation in data distribution among different authors, both in terms of their lifetime publication counts and the volume of posthumous public response, a direct visual representation would lead to significant perceptual imbalance. Thus, we applied a two-stage strategy involving Z-score standardization followed by nonlinear transformation using the Sigmoid function. This approach first removes disparities in scale across authors and then compresses extreme values, allowing for a more balanced and interpretable comparison within a unified visual framework. First, Z-score standardization was applied to eliminate dimensional inconsistencies across variables by converting raw data into a distribution with a mean of zero and a standard deviation of one. Subsequently, the standardized data were compressed using the Sigmoid function:
\begin{equation}
f(x) = \frac{1}{1 + e^{-k \cdot x}}
\end{equation}
This transformation mitigates the influence of outliers and preserves critical internal structural features. For authorial productivity data, we set the compression coefficient k=0.5 to enhance mid-range resolution and highlight subtle differences. For public attention metrics, which are more volatile, a slightly higher compression coefficient k=0.7 was applied to maintain resolution in the dense middle region while controlling expansion at the extremes. This differential parameterization facilitates a balanced and coherent visual comparison across data categories.

\subsubsection{Natural Language Processing}
To identify the lexical focus of commemorative discourse, we applied natural language processing (NLP) techniques aimed at extracting representative feature words. The analysis proceeded in two stages:

\paragraph{TF-IDF-Based Feature Extraction}

We computed Term Frequency–Inverse Document Frequency (TF-IDF) scores to identify terms with high informational value across the corpus \cite{paltoglou2010study}. These high-weighted terms reflect the unique lexical salience of each author’s commemorative narrative and serve as key indicators for further content interpretation.

\paragraph{Word Frequency Analysis and Visualization}

After standard preprocessing, including tokenization and stop-word removal, we performed frequency analysis to determine the most frequently occurring terms. The results were visualized using word clouds to reveal dominant lexical patterns, offering a concise representation of the collective linguistic attention.

\subsubsection{Semantic Foundations for Visual Construction}
\begin{figure}[tb]
 \centering 
 \includegraphics[width=\columnwidth]{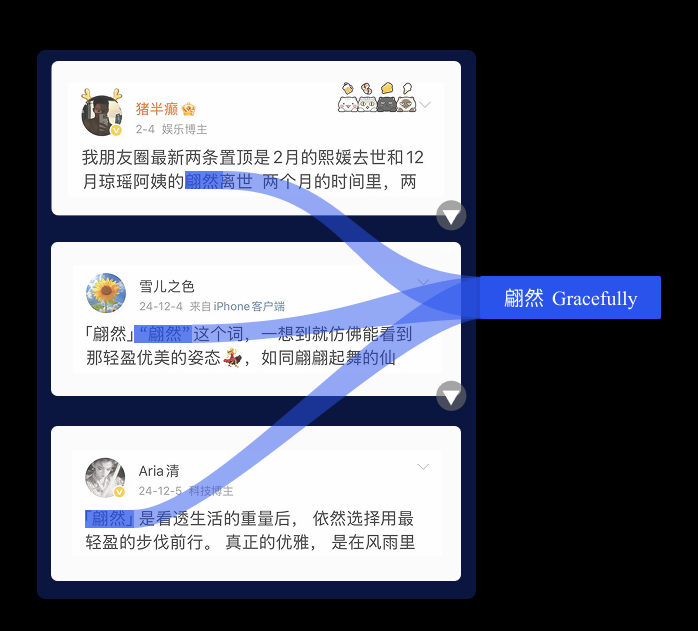}
 \caption{Posts about the theme of Gracefully (Chinese: Pianran)}
 \label{fig:3.1.2}
 
\end{figure}

Based on the aforementioned NLP pipeline, we extracted a core set of representative keywords for each author. These keywords encapsulate the collective memory associated with their life, work, and cultural presence, as constructed by the public through commemorative discourse (see \cref{fig:3.1.2}). In the subsequent stages of this artwork, these terms serve as the “semantic particles” embedded in the digital monument installation. Through visual mechanisms such as light dots, flowing text, and layered overlays, these words are spatially materialized to construct a personalized textual topography for each author—embodying a form of digital epigraphy where “words become monument.”

\subsection{Monument Visualization Design}
To translate multidimensional authorial and commemorative data into an intelligible and evocative visual form, we developed a bifurcated monument structure composed of two distinct yet interrelated segments: the lower section reflects the author’s lifetime creative output, while the upper section visualizes public response following their passing. The visual metaphor underpinning the monument design draws inspiration from the form of traditional monuments and tombstones. To foreground the core theme—that authors shape vivid characters through words in life, and are remembered through words in death—we designed a symmetrical, bifurcated structure, with the midpoint marking the transition in the author’s life course: the moment of passing. The lower segment represents the cumulative creative achievements during the author’s lifetime. Its vertical growth reflects the continual production of literary works and characters, paying tribute to the author’s own creative legacy. The upper segment, in contrast, embodies the accumulation of public commemoration after death. Here, growth is driven not by the author’s own writing, but by the collective generation of textual tributes—shared memories, reflections, and condolences—produced in online mourning. This mirrored arrangement creates a complete social portrait of the author, where personal creation and collective remembrance respond to each other. It also introduces a temporal dynamic into the memorial: while the lower section remains fixed after death, the upper section can continue to grow as long as public remembrance persists in the digital sphere. This interplay between static memory and evolving memory offers a holistic, time-sensitive form of commemoration, transforming multidimensional authorial and commemorative data into an intelligible and evocative visual form.

\subsubsection{Lower Segment: Representative works of authors}
\begin{figure}[tb]
 \centering 
 \includegraphics[width=\columnwidth]{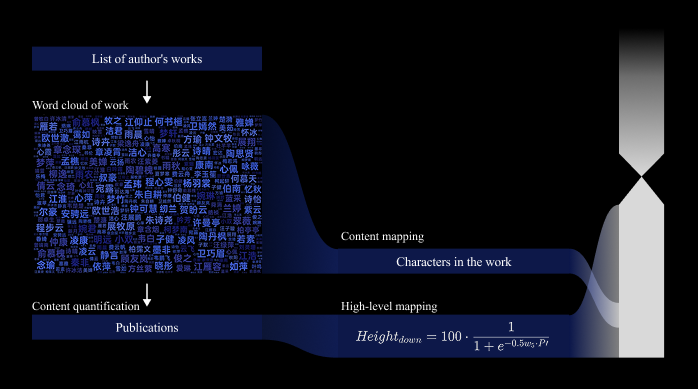}
 \caption{The composition of the lower half of the monument}
 \label{fig:3.2.2.1}
 
\end{figure}

The lower segment of each monument encodes the author’s cumulative lifetime output. Vertical height is determined by the total number of published works, processed through Z-score standardization followed by a nonlinear transformation using a Sigmoid function: 
\begin{equation}
Height{_{lower}} = 100 \cdot \frac{1}{1 + e^{-0.5 \omega_1 \cdot P}}
\end{equation}
This dual-stage transformation ensures comparability across authors with varied publication scales while preserving relational integrity. In terms of surface content, representative feature words were extracted from the authors’ own texts using TF-IDF weighting. These lexical units serve as semantic imprints of the author’s corpus, anchoring the monument’s physical form in the author’s own creative language (see \cref{fig:3.2.2.1}).

\subsubsection{Upper Segment: Posthumous Public Response}
\begin{figure}[tb]
 \centering 
 \includegraphics[width=\columnwidth]{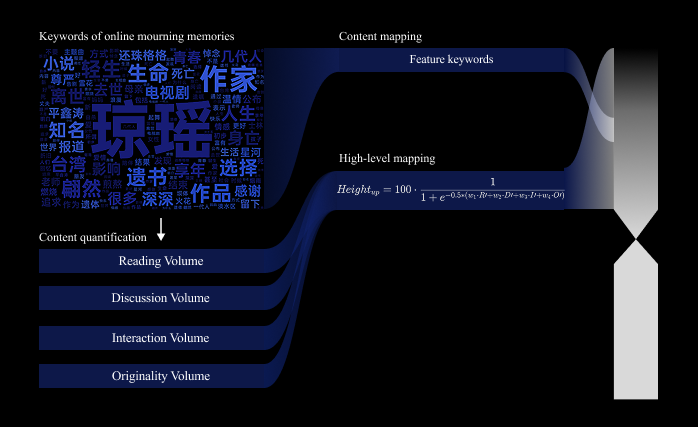}
 \caption{The composition of the upper half of the monument}
 \label{fig:3.2.1}
 
\end{figure}

The upper segment captures the intensity and nature of digital commemorative behavior following the author’s death. Height in this section is computed from four dimensions of online interaction data—Reading Volume, Discussion Volume, Interaction Volume, and Originality Volume—each subjected to Z-score normalization and sigmoid-based compression to manage scale disparities and suppress outlier distortion:
\begin{equation}
Height{_{upper}} = 100 \cdot \frac{1}{1 + e^{-0.5 \cdot (\omega_2 \cdot R + \omega_3 \cdot D + \omega_4 \cdot I + \omega_5 \cdot O)}}
\end{equation}
The textual content inscribed in this segment is derived from commemorative comments collected on social media platforms. Feature words were extracted via TF-IDF analysis to reflect key themes and emotional vocabularies present in public remembrance. As such, the upper segment functions as a real-time echo of collective memory, juxtaposed above the enduring weight of the author’s own representative works (see \cref{fig:3.2.1}).

\subsubsection{Implementation Across Selected Authors}
\begin{figure*}[t]  
    \centering
    \includegraphics[width=\textwidth]{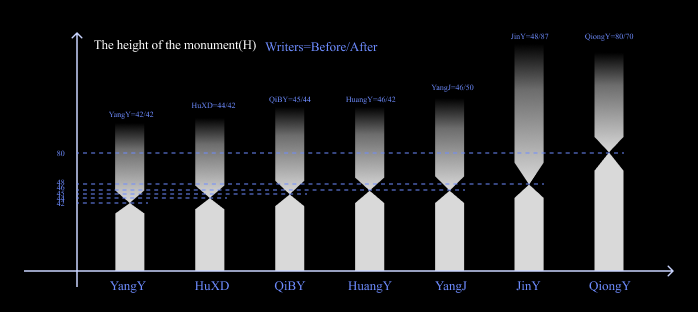} 
    \caption{The height of the monument to the seven authors}
    \label{fig:3.2.2.2}
\end{figure*}

Following this two-tiered visualization logic, we constructed individualized digital monuments for all seven selected authors (see \cref{fig:3.2.2.2}). Each monument embodies a unique integration of biographical authorship data and collective public memory, thereby transforming abstract metrics into a coherent spatial narrative. 

\begin{figure}[tb]
 \centering 
 \includegraphics[width=\columnwidth]{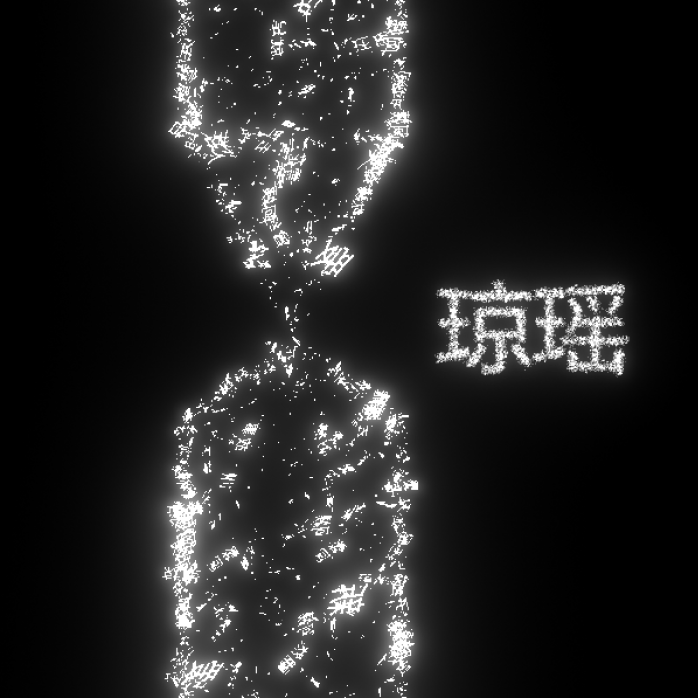}
 \caption{The static state of the monument}
 \label{fig:3.3.2}
 
\end{figure}

\begin{figure}[tb]
 \centering 
 \includegraphics[width=\columnwidth]{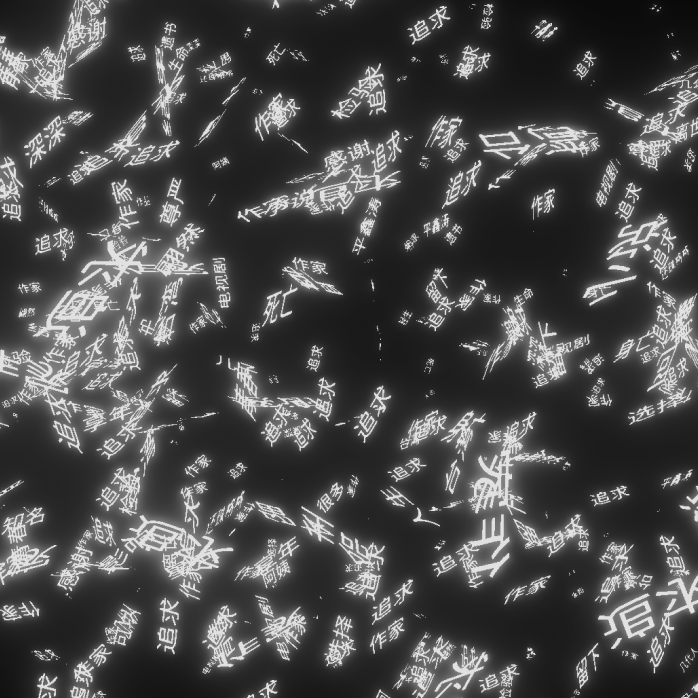}
 \caption{The dynamic state of the monument}
 \label{fig:3.3.1}
 
\end{figure}

\subsection{Interaction Design}
\begin{figure}[tb]
 \centering 
 \includegraphics[width=\columnwidth]{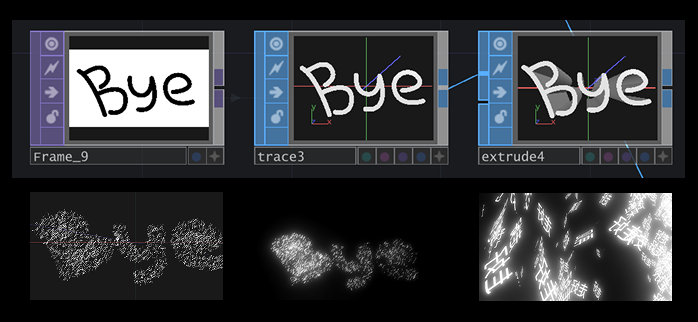}
 \caption{The process of transforming user written text into particle aggregation}
 \label{fig:3.3.3}
 
\end{figure}

The interaction design of this artwork follows a sequential user journey that transitions from initial observation to active participation. This journey is intended to gradually deepen engagement, guiding participants from visual immersion to textual exploration and finally to personal contribution. The interactive process consists of the following three stages:

\begin{itemize}
\item Immersive appreciation of the monuments: In the exploration phase, users can observe the form of each memorial and the keywords attached to it, allowing them to understand how different authors are perceived by the public, their associated characteristics, and their impact. Each memorial alternates between two states: in its non-interactive state, it appears as a static column with subtly pulsating text, symbolizing the active nature of memory (see \cref{fig:3.3.3}); when activated, it transforms into a fragmented form (see \cref{fig:3.3.1}), with keywords dispersing and moving intensely to represent the surge of discussions and significant impact triggered by the author’s passing. Pressing the button again reconstitutes the dispersed words into the static memorial. Through this dynamic, the design conveys that each memorial is constructed from fragmented pieces of information, while the dispersion of text visually represents the wave of online discussion surrounding the topic, enabling users to experience the intense emotional fluctuations within the online community.

\item Exploration of high-frequency words and original posts: By selecting high-frequency keywords, users can access related original Weibo posts (see \cref{fig:HighFrequencyWords}). This enables them to see directly how netizens express grief, share memories, and participate in online commemoration, thereby connecting the abstract visual form with concrete expressions from the public discourse.

\item Composition of personal tributes: After appreciating the memorials and exploring the original posts, users can write down their personal mourning keywords, such as ‘Bye,’ ‘Name of deceased author,’ or any word that represents their feelings (see \cref{fig:3.3.3}). Upon submission, the message is processed by a built-in multimodal AI agent, which interprets the text from the user’s input image and checks it against social media publishing rules \cite{weiboCommunityRules}. Only approved content is appended to the author’s keyword dataset. Based on the user’s input, the system uses semantic recognition and topic classification to intelligently match and present similar mourning themes from other netizens.This feature not only provides users with a space to externalize their emotions after the artistic experience, but also shows them that countless others share similar feelings and memories, thereby enhancing their sense of collective memory and fostering deeper emotional resonance.

\begin{figure}[tb]
 \centering 
 \includegraphics[width=\columnwidth]{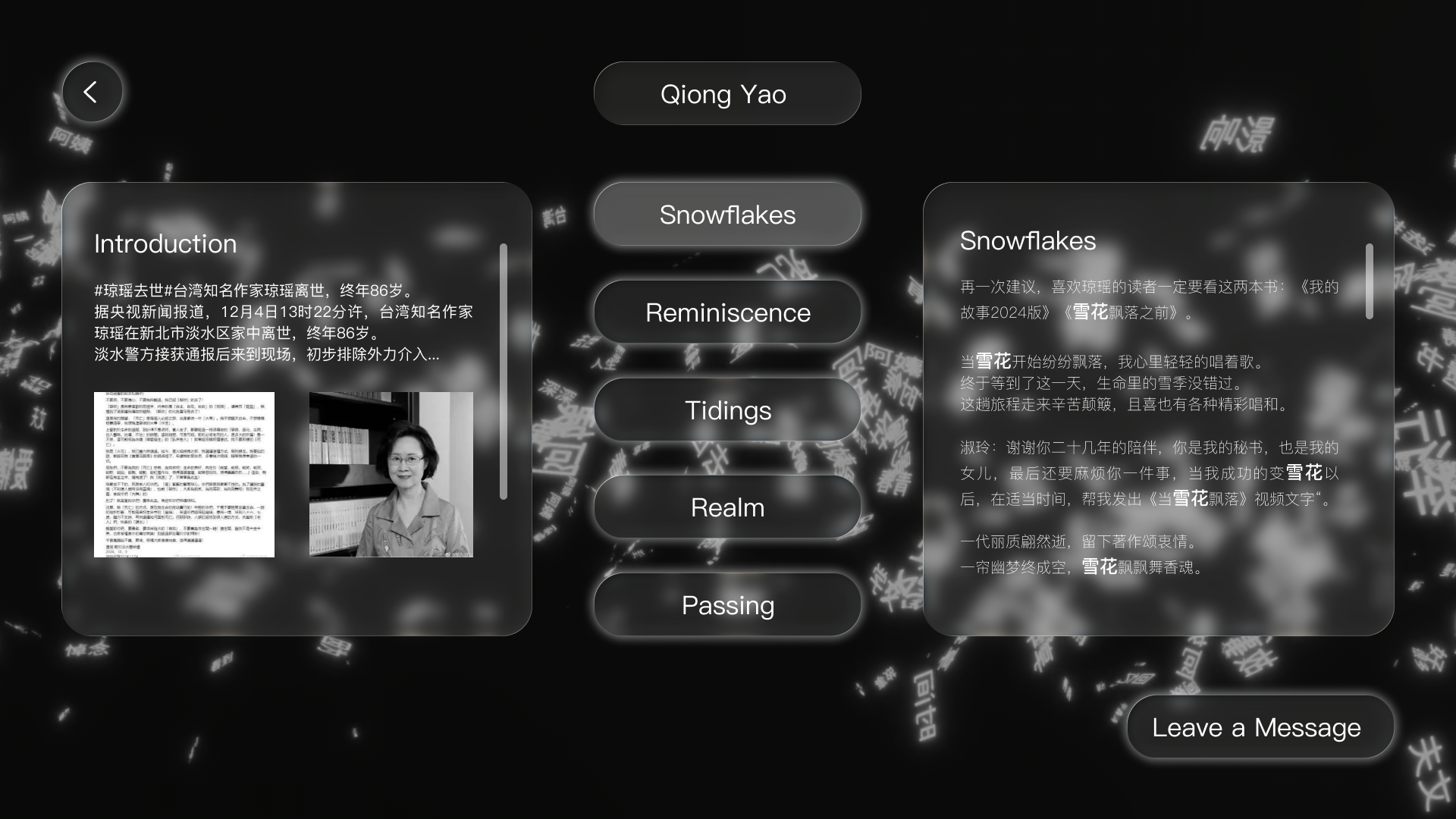}
 \caption{Original post reading interface}
 \label{fig:HighFrequencyWords}
 
\end{figure}

\end{itemize}

\begin{figure}[tb]
 \centering 
 \includegraphics[width=\columnwidth]{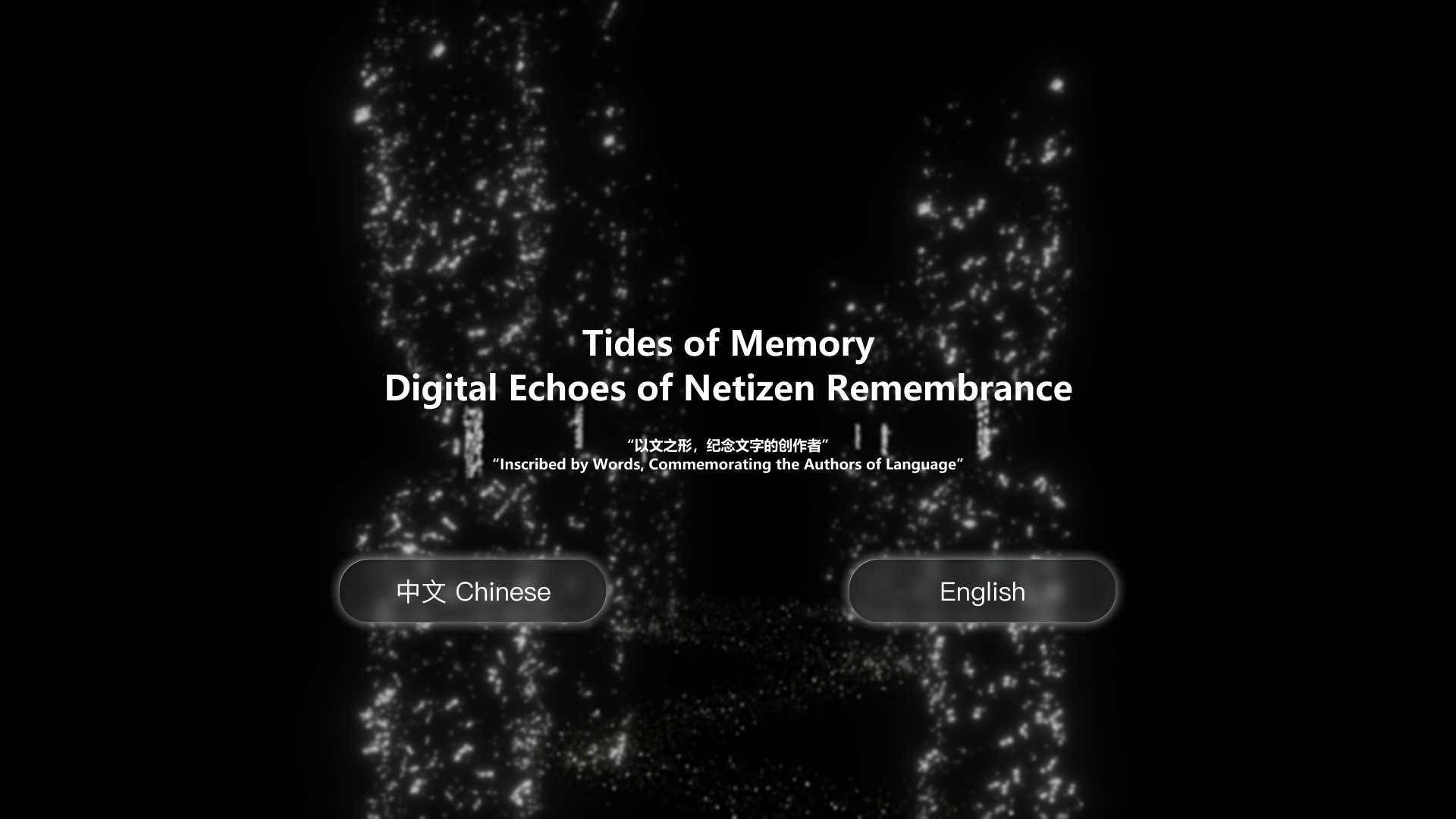}
 \caption{Language selection interface}
 \label{fig:LanguageSelection}
 
\end{figure}

\begin{figure}[tb]
 \centering 
 \includegraphics[width=\columnwidth]{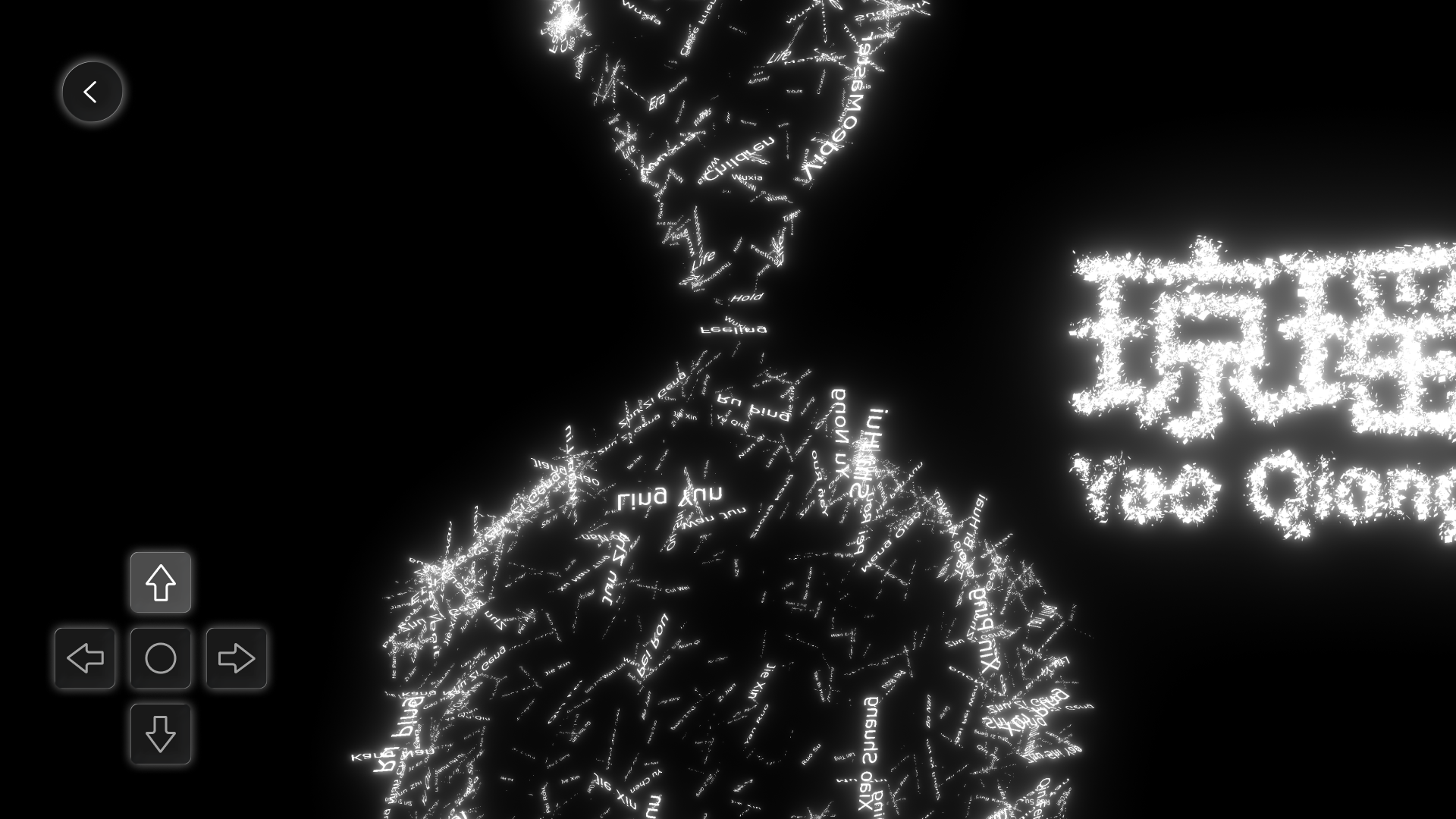}
 \caption{English version of the monument}
 \label{fig:Jump}
 
\end{figure}

To provide a more inclusive interactive experience, we implemented a bilingual interaction system in both Chinese and English, allowing participants to select their preferred language at the start of the experience (see \cref{fig:LanguageSelection}). In the English version, all high-frequency words extracted from the commemorative texts were translated into English, enabling non-Chinese speakers to access and appreciate the cultural and literary significance of these Chinese authors (see \cref{fig:Jump}).

\subsection{Interaction Scene Design}
\begin{figure}[tb]
 \centering 
 \includegraphics[width=\columnwidth]{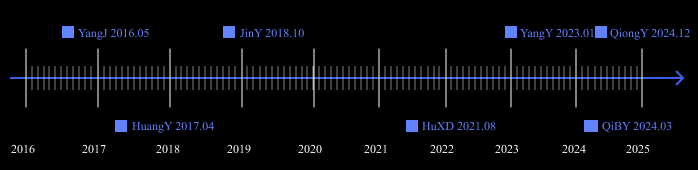}
 \caption{The dynamic state of the monument}
 \label{fig:3.4.1}
 
\end{figure}

\begin{figure}[tb]
 \centering 
 \includegraphics[width=\columnwidth]{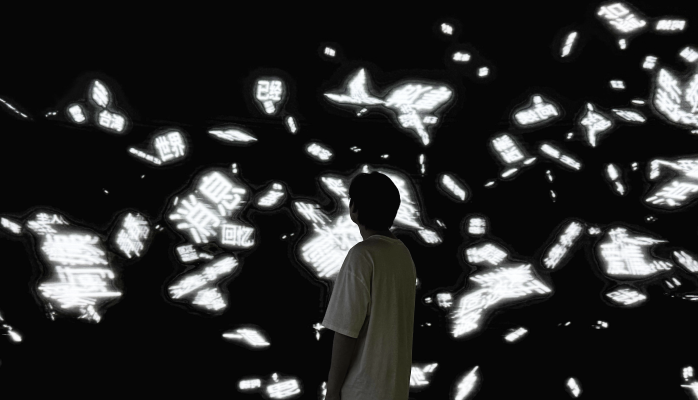}
 \caption{The effect that users appreciate in front of CAVE}
 \label{fig:3.4.2}
 
\end{figure}

In this artwork, to create an immersive experience, we placed the entire visual presentation along a Path of Remembrance (see \cref{fig:3.4.1}). It is like a hall of fame or an art corridor. Overall, the "Path of Remembrance" stretches endlessly. On both sides of the path stand monuments with authors, which are arranged according to the time of their passing, forming a clear and continuous timeline. This arrangement allows users to follow a straightforward and clear sequence of experience. The space, which is dimly lit, evokes a solemn and tranquil atmosphere. While primarily designed as a virtual experience, the scene can also be displayed on large-scale physical screens, thereby enhancing the tangibility and perceived authenticity of online mourning rituals (see \cref{fig:3.4.2}).

\begin{figure*}[t]  
    \centering
    \includegraphics[width=\textwidth]{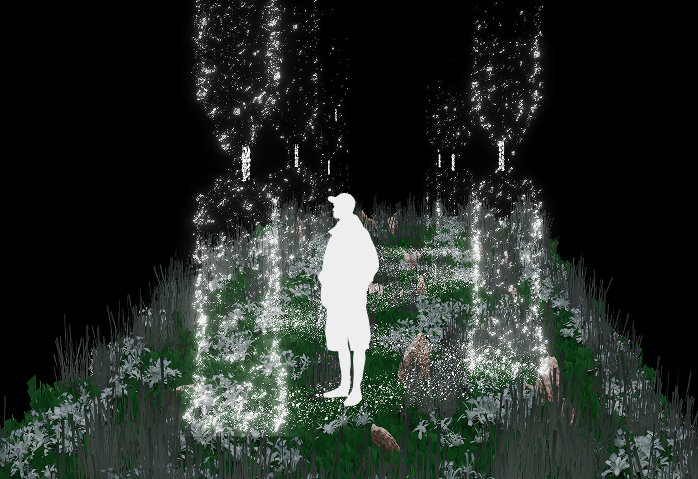} 
    \caption{3D Rendering of \textit{Tides of Memory}}
    \label{fig:3.4.3}
\end{figure*}

To further enhance and enrich the immersive user experience, this installation also supports the use of wearable virtual reality (VR) devices, enabling users to freely walk and interact within the virtual memorial space. In this digital environment, users can closely observe clusters of monuments distributed along both sides of the path, while the integrated landscape elements—such as plants—contribute to a stronger sense of immersion and environmental realism (see \cref{fig:3.4.3}). For a more profound experience, users can even enter the interior of the monuments to explore the keywords of collective memory embedded within, enabling a journey from macro to micro perspectives. This multi-layered, multi-sensory interaction design allows users to engage more deeply with emotional and cognitive aspects of digital commemoration. Moreover, with the aid of handheld controllers, users can transform their textual input from 2D surfaces into 3D models, offering a more dynamic visual effect and expanding the dimensions of exploration.

\subsection{System Development}
The visualization component of the installation was developed in TouchDesigner. Following data analysis, 3D monument models for each author were created in Rhinoceros 3D and imported into TouchDesigner. Each monument is decomposed into point clouds of varying densities according to the analytical results. High-frequency keywords for each author are stored locally as .txt files, from which the system randomly instantiates keywords onto the points during rendering. Both point density and the size of word instances can be adjusted to accommodate exhibition conditions, ensuring optimal readability.Based on our observations, most mourning hashtags exhibit limited change in daily activity, with significant fluctuations occurring primarily at the time of death and on subsequent anniversaries. Consequently, monument height changes are not immediately perceptible in real time. To address this, the design team performs periodic manual updates to incorporate new data into the visualization.

The visual dynamics are designed to convey the fragmentation of online information and the explosive growth of digital mourning in the news cycle. This is achieved by manipulating particle velocity and switching between aggregation and dispersion states. In this stage, the primary intent is to immerse the audience in the emotional and aesthetic atmosphere, rather than to present detailed textual information. To support audiences seeking closer engagement, we created a slower, more deliberate Tides of Memory mode and single-monument views. Users can navigate these views using input devices to adjust the camera perspective for closer observation.

The system also provides pathways to access original mourning posts. Users can revisit the original Weibo hashtags to view posts in their native context, recreating the social media environment at the time of the author’s passing. Additionally, a dedicated interface allows exploration of high-frequency keywords and the highly circulated posts associated with them. Keywords are selected based on frequency, while posts are curated through a combination of like and repost counts and manual filtering to ensure relevance and emotional impact.

Interaction and navigation within the installation rely on camera perspectives and scene transitions implemented in TouchDesigner. These functions can be controlled through a variety of devices—including mouse, keyboard, and gamepad—by mapping device inputs to camera movements and scene-switching actions. This flexible input mapping accommodates diverse exhibition setups and hardware configurations. For example, the keyboard may control camera motion, while the mouse can be used for scene changes, element selection, and text entry.The installation can be experienced in two hardware configurations: a large-screen mode, using a CVAE large-format display or a monitor of at least 85 inches operated via a game controller or keyboard; and a virtual reality mode, using a VR headset and controllers (e.g., HTC Vive, Oculus Quest 2) for full immersion. For optimal presentation, it is installed in a dimly lit, semi-enclosed environment with a minimum usable area of 4 × 4 meters to allow sufficient freedom of movement and maintain a high level of visual immersion.

\section{Conclusion}
\subsection{Reconstructing Collective Memory in the Digital Age}

This paper aims to explore the mechanisms of collective memory construction and emotional resonance in the digital age. By transforming a vast array of online mourning texts into dynamic digital monuments, it overcomes the physical limitations and static nature of traditional commemorative forms. This artwork highlights how personal memories reappear and circulate in public space. It encourages viewers to reflect on their place within a shared emotional network through empathetic engagement. This innovative form of memorialization brings renewed ritual meaning to digital mourning. It gives virtual memory a sense of symbolic “eternity.” The artwork also creates a closed-loop structure:“text–creator–text.” This structure reveals the power of digital media in shaping literary memory.  Ultimately, the artwork offers a poetic tribute to the deceased authors. And it contributes to the ongoing reproduction of cultural memory in contemporary digital culture.

\subsection{From Personal Grief to Collective Resonance}

People post mourning messages online, whether to express their own sorrow, share memories of the deceased, or offer condolences to the relatives and friends of the departed, are collectively shouldering and sharing the emotions of grief. The artwork focuses on individuals who participate in online mourning. It utilizes visualization to transform personal memories and emotions into perceptible forms of commemoration. It constructs a shared memorial space where each participant can perceive themselves not as isolated but as part of a collective emotional network, thereby fostering broader resonance and social cohesion. Moreover, the continuous accumulation of digital memorials allows these collective emotions to persist over time, enabling future visitors to feel a sense of connection and empathy rather than solitude—sustained by the care and emotions contributed by earlier participants. Beyond the present focus on historical literary figures, this framework could also be adapted to contexts such as the medical humanities, where archiving end-of-life narratives and expressions of condolence may serve as forms of care.

\subsection{From Data to Collective Memory}

The artwork contributes to the field of visualization by turning abstract emotions and collective memory into tangible and dynamic visual forms. It goes beyond the traditional boundaries of data visualization. It innovatively integrates the phenomenon of online mourning from the field of communication and media fields with computer science and technology. Through this integration, it addresses the formal constraints of social media platforms. It transforms the linear and 2D flow of mourning posts as a multidimensional, immersive 3D artistic space.  This shift provides participants in digital mourning with new ways to observe, reflect, and emotionally engage.
Moreover, by promoting interaction, the artwork encourages broader participation in online mourning rituals, enabling bidirectional emotional interaction and resonance among users. It transforms conventional, one-way information display into a powerful tool for fostering collective identification and the reproduction of cultural memory. The artwork extends the scope of visualization into the realms of social culture, public emotion, and digital humanities. It also shows the powerful role that visualization can play in recording, preserving, and shaping shared human experience.

\section{Limitations}
Due to the vast scale of online data, this artwork adopts a limited scope. It focuses on mourning practices related to seven Chinese-language authors, with Chinese as the primary language for analysis. Following the artistic creation methodology, artistic data visualization can focus on aesthetic and affective impact rather than measurable analytic performance, aiming to evoke emotional resonance and critical reflection through visual representation \cite{lan2025more}. Thus, no follow-up experiments were conducted to quantitatively evaluate the effectiveness of information acquisition from the visualization. In future iterations, audience evaluation will be incorporated to better understand how viewers engage with and interpret the work, complementing the current focus on artistic intent and affective impact.

In addition, while this study focuses on proposing a visualization design approach for online mourning practices, the dimensions and scale of interaction data vary considerably across different social media platforms. Applying this framework to authors from other countries or regions would therefore require recalculating the formulas according to the selected platform and the scope of authors under investigation

At the technical level, certain limitations exist. The fixed viewing angles and screen resolution of the software can affect the display of the digital artwork. As a result, the visual presentation may vary across different devices. These differences may lead to incomplete or inconsistent rendering of scenes on some platforms.

\bibliographystyle{abbrv-doi}

\bibliography{template}
\end{document}